%% file: plb_d0dplifetimes.tex
\documentclass{elsart}

\input epsf

\begin{document}

\begin{frontmatter}

\title{New measurements of the $D^0$ and $D^+$ lifetimes}
%
%
\input{author_list_plb}
%
%
\begin{abstract}
 A high statistics sample of photoproduced charm particles from
the FOCUS (E831) experiment at Fermilab has been used to measure the
$D^0$ and $D^+$ lifetimes. Using about $210\,000~D^0$ and $110\,000~D^+$
events we obtained the following values: 
$409.6 \pm 1.1~\textrm{(statistical)} \pm 1.5$~(systematic)~fs for $D^0$ and 
$1039.4 \pm 4.3~\textrm{(statistical)} \pm 7.0$~(systematic)~fs for $D^+$.
\end{abstract}

\end{frontmatter}
%


 The study of the charm hadron lifetimes has been fundamental for our
understanding of the heavy quark decays. The most important contribution is the 
spectator diagram which contributes equally to the widths of all 
hadrons of a given flavour~\cite{Bigi_Pisa}. In the early days of
charm physics it was quite a surprise when the experiments measured
a large value for $\tau_{D^+}/\tau_{D^0}$. It is generally believed that
this large ratio ($\sim 2.5$) is mainly due to the destructive interferences
between different quark diagrams that contribute only to $D^+$ decays.
The increasingly precise measurements of the heavy quark lifetimes have
stimulated the development of theoretical models, like the Heavy Quark
Theory~\cite{HQE}, which are able to predict successfully the rich pattern
of charm hadron lifetimes, that span one order of magnitude from the
longest lived ($D^+$) to the shortest lived ($\Omega^0_c$).

 In this letter we present the most accurate measurement to date 
of the lifetimes of the $D^+$ and $D^0$. Although the accuracy reached 
by the previous experiments is remarkable, for example the $D^0$ lifetime 
is known with an uncertainty of $\sim 1\%$~\cite{pdg}, we think that a more
precise determination of the $D^0$ lifetime would be needed. For example
it would allow a more accurate check for the determination of the lifetime
difference in the neutral $D$-meson system (to evaluate the parameter 
$y=\Delta\Gamma/2\Gamma$ of the $D^0-\overline{D}^0$ mixing~\cite{Jim}). 

 Charmed particles were produced by the interaction of high energy photons, obtained by 
means of bremsstrahlung of electron and positron beams (with typically $300$ GeV
endpoint energy), with a beryllium oxide target. The mean energy of the photon beam
was approximately $180$ GeV. The data were collected at Fermilab during the 
1996--97 fixed-target run. More than $6.3 \times 10^9$ triggers were collected 
from which more than 1 million charmed particles have been reconstructed.
 
 The particles from the interaction are detected in a large-aperture magnetic spectrometer with 
excellent vertex measurement, particle identification and calorimetric capabilities. The vertex
detector consists of two systems of silicon microvertex detectors. The upstream system consists
of 4 planes interleaved with the experimental target~\cite{WJohns} (2 target slab upstream then 
2 silicon planes and the replica of this setting), while the downstream system 
consists of 12 planes of microstrips arranged in three views. These detectors provide high 
resolution separation of primary (production) and secondary (decay) vertices with an average proper 
time resolution of $\sim 35~\mathrm{fs}$. The momentum of the charged particles is determined by 
measuring their deflections in two analysis magnets of opposite polarity with five stations of 
multiwire proportional chambers. Kaons and pions in the $D$-meson final states are well separated 
up to $60~{\rm GeV}/c$ of momentum using three multicell threshold \v{C}erenkov counters.


 The final states are selected using a candidate driven vertex algorithm~\cite{spectro}.
A secondary vertex is formed from the reconstructed tracks and the 
momentum vector of the $D$ candidate is used as a {\it seed} to intersect the 
other tracks in the event to find the primary vertex. Once the 
production and decay vertices are determined, the distance $\ell$ between 
them and the relative error $\sigma_\ell$ are computed. Cuts on the
$\ell/\sigma_\ell$ ratio are applied to extract the $D$ signals from the 
prompt background. The primary and secondary vertex are required to have a confidence 
level greater than $1\%$.

 The vertices (primary and secondary) have to lie inside a {\it fiducial volume}\footnote
{The reason for this cut is the presence of a trigger counter just upstream of the second microstrip device,
therefore we define the {\it fiducial volume} as the region between the first slab of the 
experimental target and this trigger counter.} and the primary vertex must be formed with 
at least two reconstructed tracks in addition to the {\it seed} track. 
The \v{C}erenkov particle identification cuts used in FOCUS are based on likelihood ratios 
between the various stable particle identification hypotheses. These likelihoods are computed 
for a given track from the observed firing response (on or off) of all cells within the 
track's ($\beta =1$) \v{C}erenkov cone for each of our three \v{C}erenkov counters. The 
product of all firing probabilities for all cells within the three \v{C}erenkov cones produces 
a $\chi^2$-like variable 
$W_i = -2 \ln (\mathrm{Likelihood})$ where $i$ ranges over the electron, pion, kaon and proton 
hypotheses (see Ref. \cite{cerenkov} for more details). We require
$\Delta_K\equiv W_\pi - W_K >1$,
called {\it kaonicity}, for the tracks reconstructed as a kaon. Analogously the tracks 
reconstructed as pions have a {\it pionicity}, $\Delta_\pi \equiv W_K - W_\pi$, exceeding 1.

 In Fig. 1a and 1b we show the invariant mass plots obtained with this set of cuts and with 
$\ell/\sigma_\ell > 9 $ for the decay modes $D^0 \to K^-\pi^+$ and $D^0 \to K^-\pi^+\pi^+\pi^-$ 
respectively (throughout this paper the charge conjugate state is implied). Fig. 1c is the
invariant mass plot for the decay mode $D^+ \to K^-\pi^+\pi^+$ with $\ell/\sigma_\ell > 14 $.
This set of cuts is chosen to optimize the yield and the background underneath 
the signal ($S/N$ ratio).

 From a binned maximum likelihood fit we find $139\,433 \pm 520 ~ D^0 \to K^-\pi^+$,
$68\,274 \pm 360 ~ D^0 \to K^-\pi^+\pi^+\pi^-$ and $109\,877 \pm 385 ~D^+ \to K^-\pi^+\pi^+$ 
candidates. The plots are fit with two Gaussians with the same mean but
different widths to take 
into account the different resolution in momentum of the tracks passing through one or two magnets
(see Ref. \cite{spectro} for more details) of our spectrometer plus a
$2^{\textrm{nd}}$ order polynomial. The low mass region is excluded in the fit to avoid possible 
contamination due to other hadronic charm decays involving an additional $\pi^0$.


 The lifetime is measured using a binned maximum likelihood fitting technique~\cite{E687_life}. 
A fit is made to the reduced proper time distribution in the signal region. The reduced proper time
is defined by $t^\prime = (\ell - N\sigma_\ell)/(\beta\gamma c)$ where $\ell$ is the distance between
the primary and the secondary vertex, $\sigma_\ell$ is the resolution on $\ell$ and $N$ is the minimum
``detachment'' cut required to extract the signal. Our vertexing algorithm provides very 
uniform reduced proper time acceptance even at very low reduced proper times. If absorption and acceptance 
corrections are small enough that they can be neglected, and if $\sigma_\ell$ is independent of $\ell$, 
one can show that the $t^\prime$ distribution for decaying charmed particles will follow an exponential 
distribution. These assumptions are very nearly true in FOCUS~\cite{spectro}. 

 The signal region reduced proper time distributions (indicated with dashed lines in Fig. 1) are formed
from events with invariant mass within $\pm 2 \overline{\sigma}$ of the mean $D$ mass; 
$\overline{\sigma}$ stands for weighted sigma because the invariant mass plots were fitted with two 
Gaussians. The dependence of the lifetime measurement on the choice of the signal and background 
region is discussed later in the text. The binned maximum likelihood method 
allows direct use of the proper time distribution of the data above and below the $D$ mass peak to 
represent the background underneath the signal instead of using a background parametrization. We 
have chosen two sidebands starting $4 \overline{\sigma}$ above and below the mean $D$ mass, each half 
as wide as the signal region (indicated with dotted lines in Fig. 1). The signal and background 
reduced proper time distributions are binned in proper time wide bins ($200$~fs) spanning about $10$ 
nominal lifetimes.

 The observed numbers of events in a reduced proper time bin $i$ (centered at $t^\prime_i$) in the 
signal and sideband histograms are labeled $s_i$ and $b_i$ respectively. The predicted number of
events $n_i$ in a reduced proper time bin is given by:
 
\begin{equation}
n_i = (N_s - B ) \frac{f(t^\prime_i)\exp (-t^\prime_i/\tau)} {\sum_i
f(t^\prime_i)\exp (-t^\prime_i/\tau)} + 
 B \frac{b_i} {\sum b_i}
\end{equation}

 where $N_s$ is the total number of events in the signal region, $B$ is the total number of background 
events in the signal region and $f(t^\prime_i)$ is a correction function. The fit parameters are $B$ and
$\tau$. The $f(t^\prime_i)$ correction function, derived from a Monte Carlo simulation, corrects the reduced
proper time evolution of the signal for the effects of geometric acceptance, reconstruction efficiency,
analysis cuts, hadronic absorption and decay of charm secondaries. The use of a multiplicative 
$f(t^\prime_i)$ correction, rather than an integral over a resolution factor, is justified since 
our reduced proper time resolution ($\sim 35~\mathrm{fs}$) is much less than the
$D^0$ or $D^+$ 
lifetime. 

 A separate $f(t^\prime)$ correction function is used for each of the three decay modes. Our Monte Carlo
simulation includes the Pythia~\cite{Pythia} model for photon-gluon fusion and incorporates a complete
simulation at the digitization level of all detector and trigger systems and
includes all known multiple
scattering and particle absorption effects. The Monte Carlo was run with $\sim 15$ times the statistics
of the experiment.

 The plots $a)$ of Fig. 2, 3 and 4 show the correction function $f(t^\prime)$ for the three decay modes 
in bins of reduced proper time. The $f(t^\prime)$ function is obtained by dividing the simulated 
reconstructed charm yield in each bin by the input decay exponential. The fall off in $f(t^\prime)$ for 
the $D^+$ case is due to the exclusion of long lived events with vertices downstream of the 
fiducial volume$^1$.

 A factor $\mathcal{L}_{\textrm{bg}}$ is included in the likelihood function in order to relate $B$ to the
number of background events expected from the side band population. The background level is thereby
jointly determined from the invariant mass distribution and from the reduced proper time evolution
in the side bands. The likelihood function is then given by: 
\begin{equation}
\mathcal{L} = \mathcal{L}_{\textrm{signal}}\times\mathcal{L}_{\textrm{bg}}
\end{equation}
 where
\begin{equation}
\mathcal{L}_{\textrm{signal}} = \prod_{i=1}^{\textrm{bins}} \frac{n_i^{s_i}\exp(-n_i)} {s_i!}
\end{equation}
 and
\begin{equation}
\mathcal{L}_{\textrm{bg}} = \frac{B^{N_{\textrm{bg}}}} {N_{\textrm{bg}}!} \exp(-B)
\end{equation} 
 with $N_{\textrm{bg}} = \sum_i b_i$ (we assume a linear background because the 
$2^{\textrm{nd}}$ order term of the polynomial is negligible).

 The plots $b)$ of Fig. 2, 3 and 4 show the predicted events (histogram)
 superimposed on
the observed events, the background events $b_i$ are also superimposed. In plots $c)$ of Fig. 2, 3 
and 4 a pure exponential function with the fitted lifetime is superimposed on the background 
subtracted and $f(t^\prime)$ corrected $t^\prime$ distribution. 

 The measured lifetimes are: $408.75 \pm 1.42$~fs for $D^0 \to K^-\pi^+$, 
$411.25 \pm 1.95$~fs for $D^0 \to K^-\pi^+\pi^+\pi^-$ and $1039.42 \pm 4.28$~fs for 
$D^+ \to K^-\pi^+\pi^+$.
 
  Our lifetime measurements have been tested by modifying each of the vertex and \v{C}erenkov
cuts individually. For example in Fig. 5 one can see the measured lifetimes versus 
the $\ell/\sigma_\ell$ detachment cut. The measured lifetimes of the three decay modes are
stable with respect to $\ell/\sigma_\ell$. Our measured lifetimes show no significant variation 
with the cuts employed to extract the signal. 

 We check our $D^0$ lifetime evaluation partitioning the total sample into
 $D^\star-$tag and \emph{no-tag}
according to their origin. The obtained lifetimes are in very good agreement with the reported values. 

 To further check our lifetime measurements we have used tight cuts in order to
 extract a signal with
virtually no background. Fig. 6 shows the invariant mass plots of the three decay modes for this
set of cuts. The lifetime measurements from these samples, $411.26 \pm 3.11$~fs for $D^0 \to K^-\pi^+$ 
$413.10 \pm 4.80$~fs for $D^0 \to K^-\pi^+\pi^+\pi^-$ and $1036.66 \pm 8.00$~fs for 
$D^+ \to K^-\pi^+\pi^+$, are in very good agreement with our previous determination.


 Systematic uncertainties for these lifetime measurements can arise from several sources. 
We performed a detailed study to analyze these sources.

 There is an uncertainty due to the absolute time scale which was determined by studying the 
absolute length and momentum scale in the experiment (see Ref. \cite{Harry} for more
details). We estimate an uncertainty of $\pm 0.11\%$ for this source.

 Another source of systematic uncertainty is linked to the detector and reconstruction efficiency.
The $f(t^\prime)$ corrects the reduced proper time distribution for these effects, but an 
uncertainty could originate if there is a mismatch between the Monte Carlo simulation and 
the data. We have verified that our Monte Carlo accurately reproduces the distributions of several
relevant variables, such as the longitudinal and transverse momenta, the multiplicity of the 
production vertex, the measured decay length and the estimated error on the reconstructed
proper time. In order to estimate this uncertainty we split our total 
sample into independent subsamples depending on $D$ momentum, particle versus anti-particle and the
different periods in which the data were collected. The splits into $D$ momentum and charge conjugation
are the natural tests to reveal a possible mismatch between data and Monte Carlo because they probe the
response of the detector. The main reason for the period dependence is the insertion of 
the upstream silicon system (which improved the resolution) in the target 
region during the 1997 fixed-target run period. A technique, employed in FOCUS 
and in the predecessor experiment E687, modeled after the {\it S-factor method} from the 
Particle Data Group~\cite{pdg}, was used to try to separate true systematic variations from statistical 
fluctuations. The lifetime is evaluated for each of the $8~(=2^3)$ statistically independent subsamples 
and a {\it scaled variance} is calculated; the {\it split sample} variance is defined as the difference
between the reported statistical variance and the scaled variance if the scaled variance 
exceeds the statistical variance. This contribution to the systematic error is reported as 
{\it split sample} in Table 1.
  
 The reported lifetimes are obtained with a particular set of fitting conditions. For example the
width of the bins or the range of the $t^\prime$ distribution. This is a particular choice and
the lifetime should be independent of it. We investigated if this could be a possible source of 
uncertainty by varying the width of the bins, the upper limit of the $t^\prime$ distribution, the location
and the width of the sidebands, and the width of the signal region. In addition we studied the 
effect of using only the low or only the high mass sideband as well as the effect of 
eliminating the background term in the likelihood (the second term in equation $2$). For
all these {\it fit variants} the sample variance is used as an estimate of this
uncertainty because the various measurements are all taken as {\it a priori} likely.

 A further source of systematic error can be due to uncertainties in the target absorption corrections.
Two effects are present: hadronic absorption of decay daughters which would increase the fitted 
lifetime if neglected and absorption of the $D$ in the target which would tend to decrease the 
fitted lifetime if not taken into account. In our Monte Carlo all known particle absorption 
effects (values from Particle Data Group~\cite{pdg}) have been simulated for the decay daughters. 
For the charm hadron absorption our simulation assumes $1/2$ of the cross section for neutrons.
We estimate the uncertainty of this contribution by varying the charm cross-section by $50\%$ and the daughter particle
interaction cross-sections by $25\%$ in the Monte Carlo simulation. We verify that these estimates,
reported in Table 1, are consistent with a determination of this contribution obtained
comparing the lifetimes of decays with the $D$ produced in the upstream half of each target with
those produced in the downstream half of the same target (see Ref. \cite{Focus_Genoa} for more
details on the target setup). Each partition represents a different mixture of
hadronic absorptions
of decay daughters and $D$ mesons. 

 The acceptance could be another source of uncertainty. We analyzed this effect determining the $D^0$ lifetime
without the correction function $f(t^\prime)$ and removing the fiducial volume cut from the set of analysis 
cuts (this makes the correction function almost flat). We obtained results in good agreement with the reported values.
This check is not possible for the $D^+$ because of the longer lifetime, a
geometric acceptance correction is always
needed. A study was performed in FOCUS (see Ref. \cite{Harry} for more details) comparing the acceptance
part of the Monte Carlo correction with the high statistics $K^0_S \to \pi^+\pi^-$ decays. The result of this
study showed an excellent agreement between the acceptance observed in the data and the acceptance simulated 
by the Monte Carlo; however we assess a $2\%$ uncertainty due to the finite statistics. This $2\%$ uncertainty
in the $f(t^\prime)$ correction function gives a $0.21\%$ and $0.52\%$ uncertainty in the lifetime of
$D^0$ and $D^+$ respectively.

  The finite Monte Carlo statistics give a negligible contribution to the 
systematic uncertainty.

\begin{table}
\caption{Contributions in percent to the systematic uncertainty.} 
\begin{tabular}{|l|cccc|}
\hline
                          &  $D^0$     &    $D^0$	      &  $D^0$      &  $D^+$	     \\
 Source                   & $K^-\pi^+$ & $K^-\pi^+\pi^+\pi^-$ &  combined   & $K^-\pi^+\pi^+$\\
\hline
 Absolute time scale      & $0.11\%$   &  $0.11\%$	      &	$0.11\%$    &	$0.11\%$     \\
 Split sample             & $0.$       &  $0.37\%$	      &	$0.13\%$    &	$0.$	     \\
 Fit variant              & $0.19\%$   &  $0.14\%$	      &	$0.17\%$    &	$0.15\%$     \\
 Absorption               & $0.11\%$   &  $0.20\%$	      &	$0.20\%$    &	$0.38\%$     \\
 Acceptance               & $0.21\%$   &  $0.21\%$	      &	$0.21\%$    &	$0.52\%$     \\
\hline
 Total systematic error   & $0.32\%$   &  $0.50\%$	      &	$0.38\%$    &	$0.67\%$     \\
\hline 
\end{tabular}
\end{table}
  
 Table 1 shows the contributions of each of these sources to the total systematic uncertainty.
For the combined $D^0$ lifetime the systematic error (also shown in Table 1) is obtained combining the 
individual sources of systematic uncertainty from $D^0 \to K^-\pi^+$ and $D^0 \to K^-\pi^+\pi^-\pi^+$.
We assume the absolute time scale and the absorption correlated. To obtain the final systematic error 
the uncertainties from the different sources are then added in quadrature.
  
 The final lifetime values are $409.62 \pm 1.15~\textrm{(statistical)} \pm 1.55$~(systematic)~fs
for $D^0$ (weighted average) and $1039.42 \pm 4.28~\textrm{(statistical)}\pm 6.97$~(systematic)~fs 
for $D^+$. A final check was performed for the $D^0$ lifetime. We compute the lifetime using a 
combined likelihood, that is forming a global likelihood for $K^-\pi^+$ and $K^-\pi^+\pi^-\pi^+$. 
The fit parameters are the two distinct backgrounds and one lifetime. The lifetime from the 
combined likelihood, $409.62 \pm 1.15$~(statistical), is identical to the reported value.

 This measurement of the $D^0$ lifetime value is in very good agreement with the result we 
obtained in our lifetime difference paper~\cite{Jim}.


 In conclusion we have measured the lifetimes of the $D^0$ and $D^+$ mesons. Our results are reported in 
Table 2 along with a comparison with the most recent published measurements. 
 
\begin{table}[ht!]
\caption{Measured lifetimes ($\times 10^{-12}$~s).}
\begin{tabular}{|l|l|l|}
\hline  
 Experiment     & $D^0$    & $D^+$                                                                     \\
\hline
  E687 \cite{E687_life}     & $0.413  \pm 0.004  \pm 0.003$            & $1.048  \pm 0.015  \pm 0.011$  \\
  CLEO II \cite{CLEO_life} & $0.4085 \pm 0.0041 ^{+0.0035}_{-0.0034}$ & $1.0336 \pm 0.0221 ^{+0.0099}_{-0.0127}$ \\
  E791 \cite{E791_life}    & $0.413  \pm 0.003  \pm 0.004$            &                                \\
  This measurement         & $0.4096 \pm 0.0011 \pm 0.0015$           & $1.0394 \pm 0.0043 \pm 0.0070$ \\
\hline  
\end{tabular}
\end{table}

 Our results will significantly decrease the errors on the current world average values for the 
$D^0$ and $D^+$ lifetimes.

 From our measurements of the $D^0$ and $D^+$ lifetimes we can update the determination of
the ratio $\tau(D^+)/\tau(D^0)$: $2.538 \pm 0.023$. This result and the inclusive 
semileptonic branching ratios~\cite{pdg}, $D^+\rightarrow eX = (17.2 \pm 1.9)\%$
and $D^0\rightarrow eX = (6.75 \pm 0.29)\%$, show that the 
$D^0$ and $D^+$ semileptonic decay widths are nearly equal:  
\begin{equation}
 \frac{\Gamma(D^0\rightarrow eX)}{\Gamma(D^+\rightarrow eX)} =
 \frac{B(D^0\rightarrow eX)}{B(D^+\rightarrow eX)} \times 
 \frac{\tau(D^+)}{\tau(D^0)} = 1.00 \pm 0.12
\end{equation}

 This implies that differences in the total decay widths between $D^0$ and
$D^+$ must be due to differences in the hadronic decay sector.
 
\vspace{1.cm}

We wish to acknowledge the assistance of the staffs of Fermi National
Accelerator Laboratory, the INFN of Italy, and the physics departments of
the
collaborating institutions. This research was supported in part by the U.~S.
National Science Foundation, the U.~S. Department of Energy, the Italian
Istituto Nazionale di Fisica Nucleare and Ministero dell'Universit\`a e
della Ricerca Scientifica e Tecnologica, the Brazilian Conselho Nacional de
Desenvolvimento Cient\'{\i}fico e Tecnol\'ogico, CONACyT-M\'exico, the
Korean Ministry of Education, and the Korea Research Foundation.
%
%
%
%
%

%
%
%
%
\begin{figure}[ht!]
\vspace{15cm}
\epsfxsize=9.4mm
\epsfbox{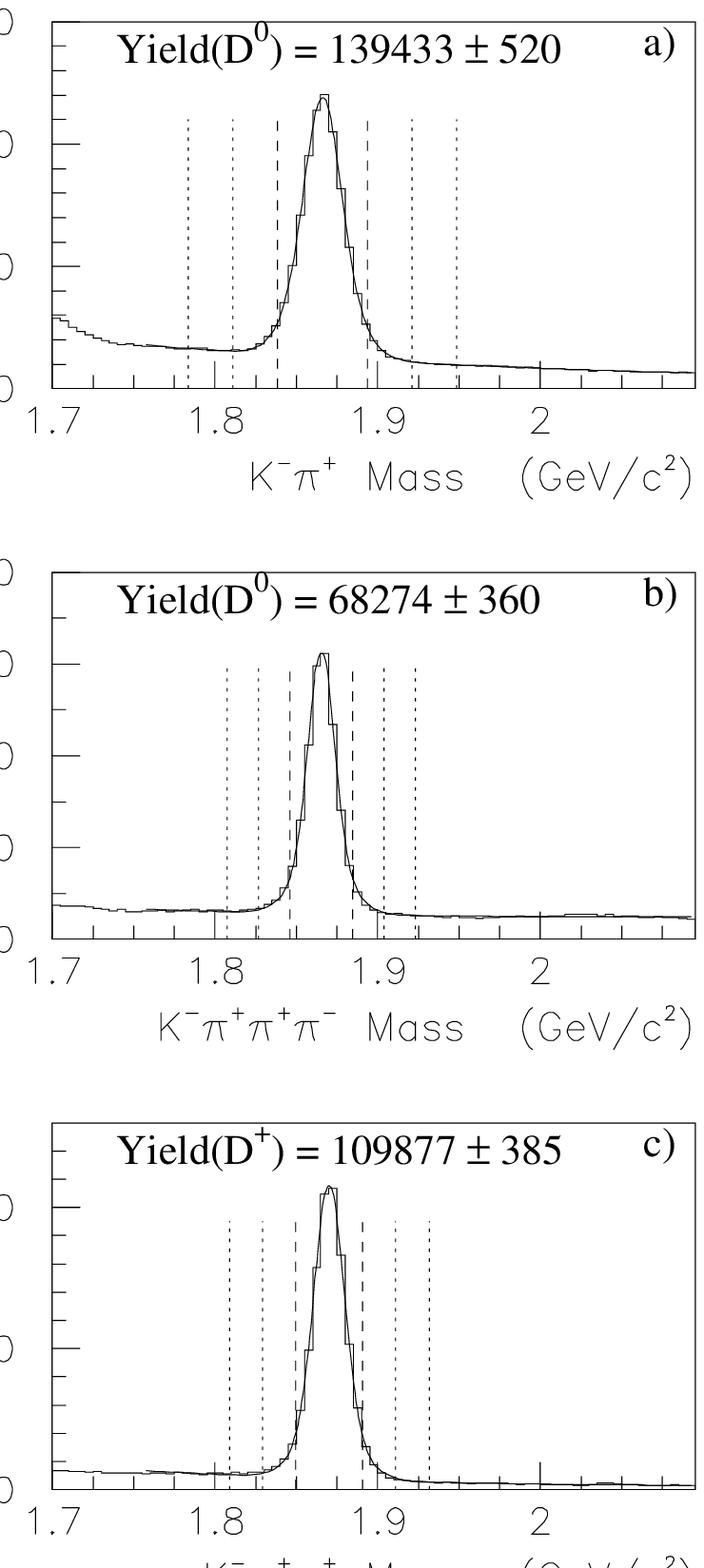}
\vspace{0.5cm} 
\caption{ (a) $K^-\pi^+$ invariant mass distribution, 
 (b) $K^-\pi^+\pi^+\pi^-$ invariant mass distribution, (c) $K^-\pi^+\pi^+$
 invariant mass distribution. The fits (solid curves) are described in 
 the text and the numbers quoted are the yields. The vertical dashed lines indicate
 the signal region, the vertical dotted lines the sideband. }
\end{figure}
\newpage
%
%
\begin{figure}[ht!]
\vspace{15cm}
\epsfxsize=9.4mm
\epsfbox{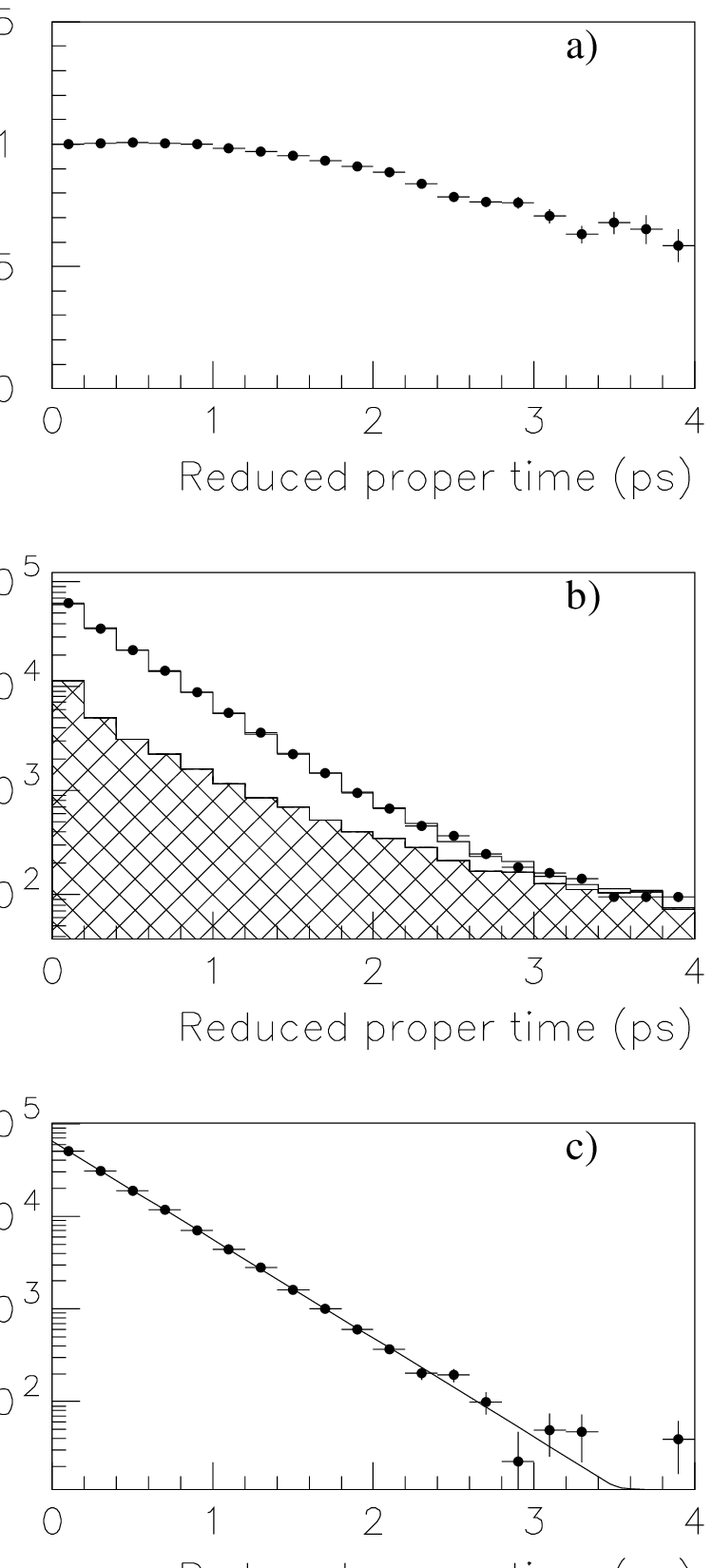}
\vspace{0.5cm} 
\caption{ Decay mode $D^0 \to K^-\pi^+$: (a) the correction function $f(t^\prime)$, deviation 
from a flat distribution represents the correction from a pure exponential function; (b) 
the predicted events (histogram) are superimposed to the observed events (points), 
the shaded distribution shows the $t^\prime$ distribution of the background;
(c) the background subtracted and $f(t^\prime)$ corrected $t^\prime$ distribution,
the superimposed straight line is an exponential with the fitted lifetime.}
\end{figure}
\newpage
%
%
\begin{figure}[ht!]
\vspace{15cm}
\epsfxsize=9.4mm
\epsfbox{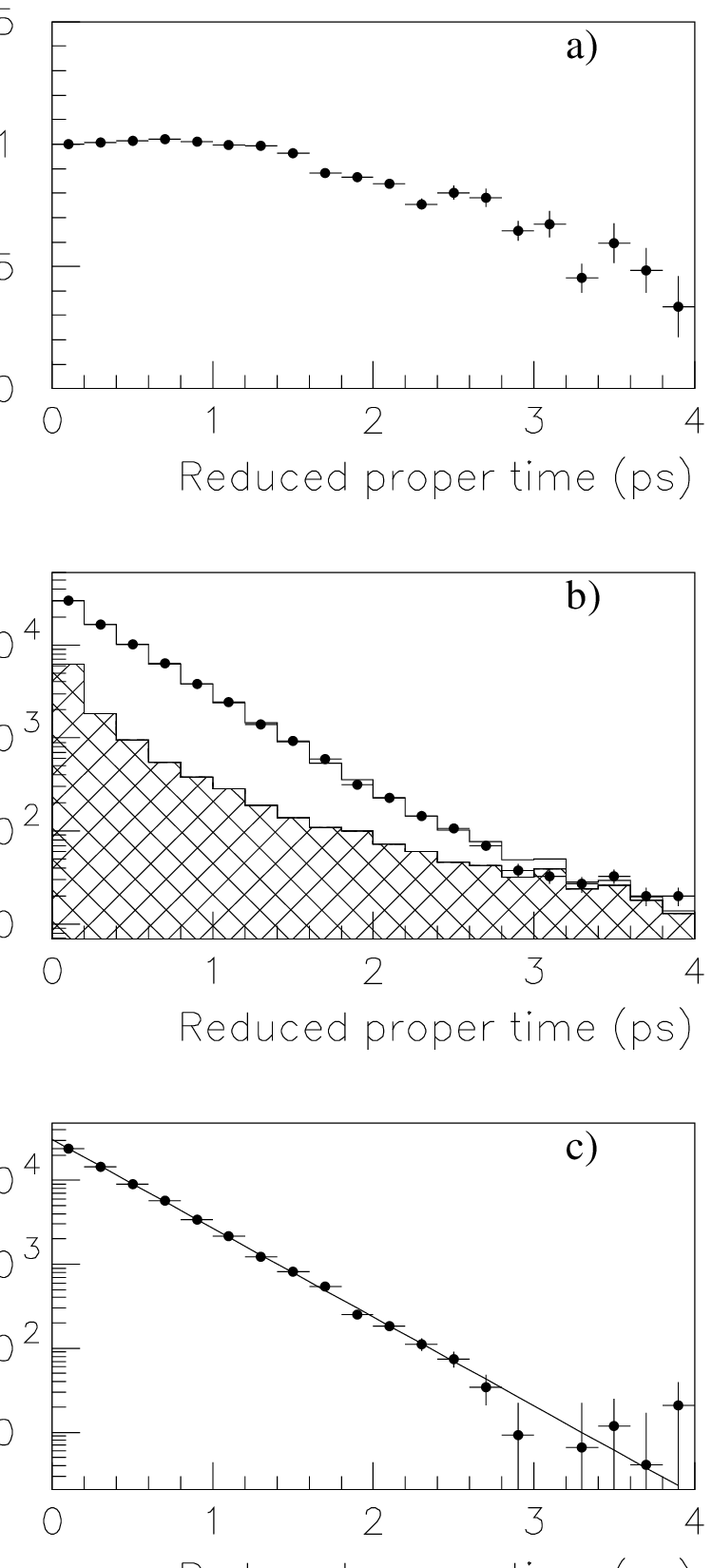}
\vspace{0.5cm} 
\caption{ Decay mode $D^0 \to K^-\pi^+\pi^-\pi^+$: (a) the correction function $f(t^\prime)$, deviation 
from a flat distribution represents the correction from a pure exponential function; (b) 
the predicted events (histogram) are superimposed to the observed events (points), 
the shaded distribution shows the $t^\prime$ distribution of the background;
(c) the background subtracted and $f(t^\prime)$ corrected $t^\prime$ distribution,
the superimposed straight line is an exponential with the fitted lifetime.}
\end{figure}
\newpage
%
%
\begin{figure}[ht!]
\vspace{15cm}
\epsfxsize=9.4mm
\epsfbox{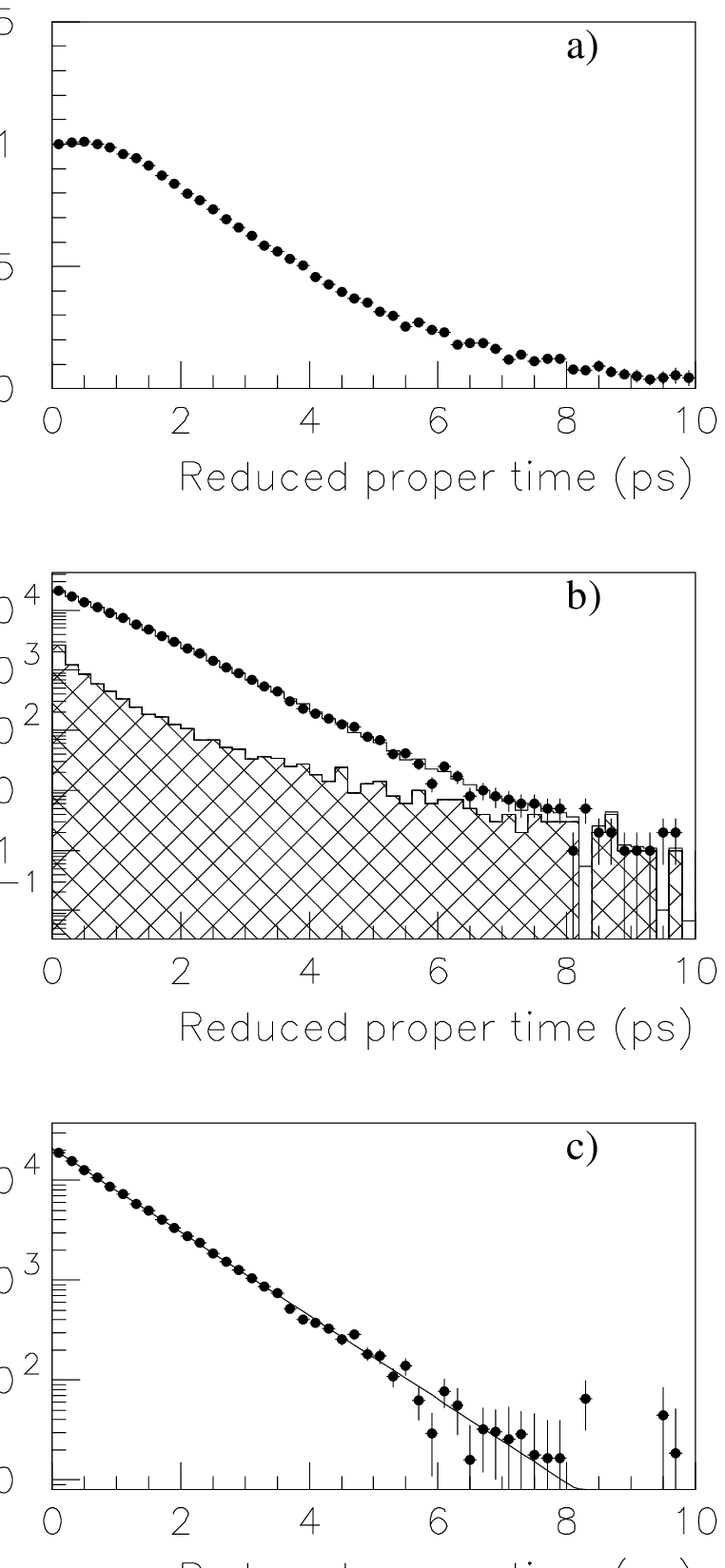}
\vspace{0.5cm} 
\caption{ Decay mode $D^+ \to K^-\pi^+\pi^+$: (a) the correction function $f(t^\prime)$, deviation 
from a flat distribution represents the correction from a pure exponential function; (b) 
the predicted events (histogram) are superimposed to the observed events (points), 
the shaded distribution shows the $t^\prime$ distribution of the background;
(c) the background subtracted and $f(t^\prime)$ corrected $t^\prime$ distribution,
the superimposed straight line is an exponential with the fitted lifetime.}
\end{figure}
\newpage
%
%
\begin{figure}[ht!]
\vspace{15cm}
\epsfxsize=9.4mm
\epsfbox{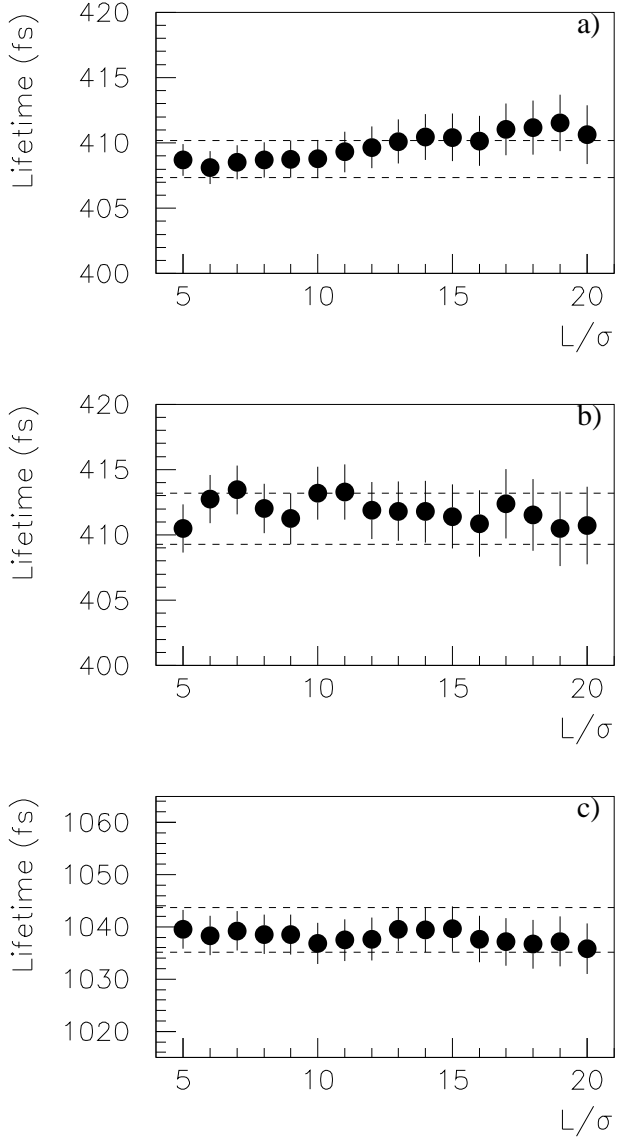}
\vspace{0.5cm} 
\caption{ The fitted lifetime versus the $\ell/\sigma_\ell$ detachment cut for:
 (a) $D^0 \to K^-\pi^+$, (b) $D^0 \to K^-\pi^+\pi^+\pi^-$ and (c) $D^+ \to K^-\pi^+\pi^+$. 
 The horizontal dashed lines show the interval corresponding to the chosen lifetime 
 $\pm 1 \sigma$. }  
\end{figure}
\newpage
%
%
\begin{figure}[ht!]
\vspace{15cm}
\epsfxsize=9.4mm
\epsfbox{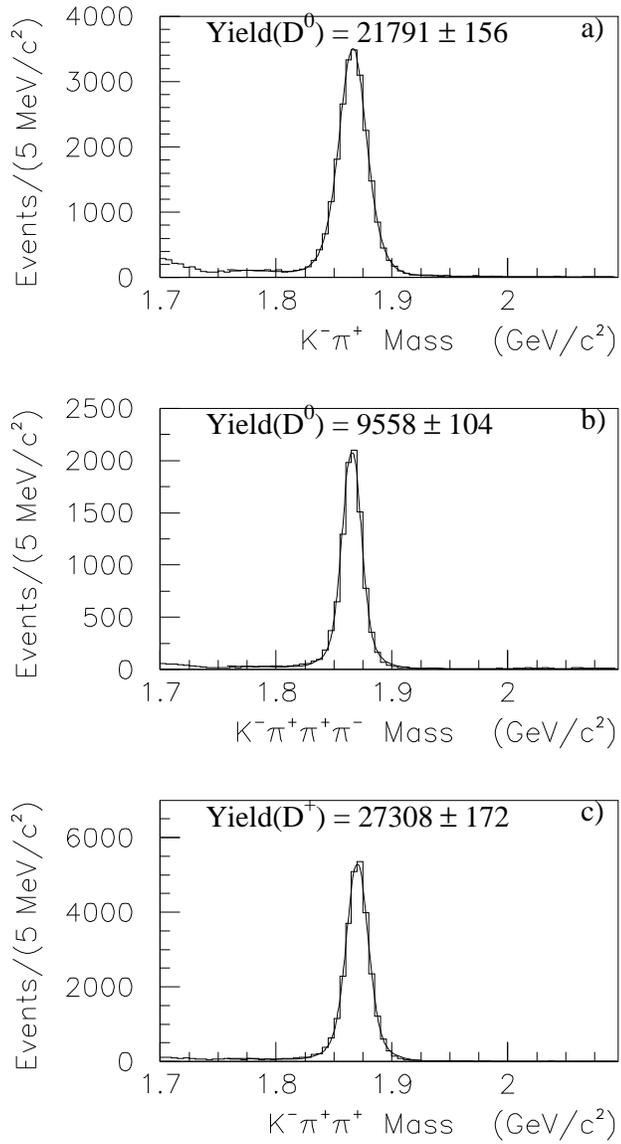}
\vspace{0.5cm} 
\caption{ Invariant mass distributions obtained using tight cuts for:
 (a) $K^-\pi^+$, (b) $K^-\pi^+\pi^+\pi^-$ and (c) $K^-\pi^+\pi^+$.
 The functions used to fit the data (solid curves) are similar to those of 
 Figure 1 and the numbers quoted are the yields. }
\end{figure}
%
\end{document}

%% file: author_list_plb.tex
$\textrm{The~FOCUS~Collaboration}^\star$
\thanks{See \textrm{http://www-focus.fnal.gov/authors.html} for
additional author information.}
\author[ucd]{J.~M.~Link},
\author[ucd]{M.~Reyes},
\author[ucd]{P.~M.~Yager},
\author[cbpf]{J.~C.~Anjos},
\author[cbpf]{I.~Bediaga},
\author[cbpf]{C.~G\"obel},
\author[cbpf]{J.~Magnin},
\author[cbpf]{A.~Massafferri},
\author[cbpf]{J.~M.~de~Miranda},
\author[cbpf]{I.~M.~Pepe},
% ...
\address[ucd]{University of California, Davis, CA 95616} \address[cbpf]{Centro 
Brasileiro de Pesquisas F\'isicas, Rio de Janeiro, RJ, Brasil} \address[cinv]{CINVESTAV, 
07000 M\'exico City, DF, Mexico} \address[cu]{University of Colorado, Boulder, CO 80309}
\address[fnal]{Fermi National Accelerator Laboratory, Batavia, IL 60510} \address[fras]
{Laboratori Nazionali di Frascati dell'INFN, Frascati, Italy I-00044} \address[ui]
{University of Illinois, Urbana-Champaign, IL 61801} \address[iu]{Indiana University, 
Bloomington, IN 47405} \address[korea]{Korea University, Seoul, Korea 136-701} \address[kp]
{Kyungpook National University, Taegu, Korea 702-701} \address[milan]{INFN and University 
of Milano, Milano, Italy} \address[nc]{University of North Carolina, Asheville, NC 28804}
\address[pavia]{Dipartimento di Fisica Nucleare e Teorica and INFN, Pavia, Italy} \address[pr]
{University of Puerto Rico, Mayaguez, PR 00681} \address[sc]{University of South Carolina, 
Columbia, SC 29208} \address[ut]{University of Tennessee, Knoxville, TN 37996} \address[vu]
{Vanderbilt University, Nashville, TN 37235} \address[wisc]{University of Wisconsin, Madison, 
WI 53706}